\documentclass[amsmath,amssymb,twocolumn,prd,floatfix,showpacs, nofootinbib]{revtex4-1}
\usepackage{tabularx}
\usepackage{bm, color}
\usepackage{overpic,subfigure}
\usepackage{multirow}
\usepackage{array}
\usepackage{dcolumn}
\usepackage[symbol]{footmisc}
\usepackage{epstopdf}
\usepackage{ulem}
\RequirePackage{xspace}

\newcommand{\gev}{\ensuremath{\mathrm{\,Ge\kern -0.1em V}}\xspace}
\newcommand{\mev}{\ensuremath{\mathrm{\,Me\kern -0.1em V}}\xspace}
\newcommand{\mevcc}{\ensuremath{{\mathrm{\,Me\kern -0.1em V\!/}c^2}}\xspace}

\newcommand{\jprBase}        {Phys.\ Rev.\xspace}
\newcommand{\jplBase}        {Phys.\ Lett.\xspace}
\newcommand{\plb}      [1]{\jplBase\ {B~#1}}
\newcommand{\jprd}     [1]{\jprBase\ {D~#1}}

\def\chic#1{\ensuremath{\chi_{c#1}}\xspace}

\def\fz#1       {\ensuremath{f_0({#1})}\xspace}

\usepackage{lineno}
\usepackage{hyperref}
\hypersetup{colorlinks = true,
            linkcolor = black,
            urlcolor = blue,
            citecolor = blue}
\begin{document}
\title{\boldmath Observation of $\chi_{cJ}\to 3(K^+K^-)$}

\author{
M.~Ablikim$^{1}$, M.~N.~Achasov$^{4,c}$, P.~Adlarson$^{75}$, O.~Afedulidis$^{3}$, X.~C.~Ai$^{80}$, R.~Aliberti$^{35}$, A.~Amoroso$^{74A,74C}$, Q.~An$^{71,58,a}$, Y.~Bai$^{57}$, O.~Bakina$^{36}$
% ...
\\
\vspace{0.2cm}
(BESIII Collaboration)\\
\vspace{0.2cm} {\it
$^{1}$ Institute of High Energy Physics, Beijing 100049, People's Republic of China\\
% ...
$^{3}$ Bochum  Ruhr-University, D-44780 Bochum, Germany\\
$^{4}$ Budker Institute of Nuclear Physics SB RAS (BINP), Novosibirsk 630090, Russia\\
% ...
$^{35}$ Johannes Gutenberg University of Mainz, Johann-Joachim-Becher-Weg 45, D-55099 Mainz, Germany\\
$^{36}$ Joint Institute for Nuclear Research, 141980 Dubna, Moscow region, Russia\\
% ...
$^{57}$ Southeast University, Nanjing 211100, People's Republic of China\\
$^{58}$ State Key Laboratory of Particle Detection and Electronics, Beijing 100049, Hefei 230026, People's Republic of China\\
% ...
$^{71}$ University of Science and Technology of China, Hefei 230026, People's Republic of China\\
% ...
$^{74}$ University of Turin and INFN, (A)University of Turin, I-10125, Turin, Italy; (B)University of Eastern Piedmont, I-15121, Alessandria, Italy; (C)INFN, I-10125, Turin, Italy\\
$^{75}$ Uppsala University, Box 516, SE-75120 Uppsala, Sweden\\
% ...
$^{80}$ Zhengzhou University, Zhengzhou 450001, People's Republic of China\\
\vspace{0.2cm}
$^{a}$ Deceased\\
% ...
$^{c}$ Also at the Novosibirsk State University, Novosibirsk, 630090, Russia\\
% ...
}
\vspace{0.4cm}
}

\begin{abstract}
  By analyzing $(27.12\pm0.14)\times10^8$ $\psi(3686)$ events collected with the BESIII detector operating at the BEPCII collider, the decay processes $\chi_{cJ} \to 3(K^+K^-)$ ($J=0,1,2$) are observed for the first time with statistical significances of 8.2$\sigma$, 8.1$\sigma$, and 12.4$\sigma$, respectively. The product branching fractions of $\psi(3686)\to\gamma\chi_{cJ}$, $\chi_{cJ}\to 3(K^+K^-)$ are presented and the branching fractions of $\chi_{cJ}\to 3(K^+K^-)$ decays are determined to be
  $\mathcal{B}_{\chi_{c0}\to 3(K^+K^-)}$=$(10.7\pm1.8\pm1.1)$$\times10^{-6}$,
  $\mathcal{B}_{\chi_{c1}\to 3(K^+K^-)}$=$(4.2\pm0.9\pm0.5)$$\times10^{-6}$, and
  $\mathcal{B}_{\chi_{c2}\to 3(K^+K^-)}$=$(7.2\pm1.1\pm0.8)$$\times10^{-6}$,
  where the first uncertainties are statistical and the second are systematic.

\end{abstract}

\maketitle
%\linenumbers

\section{Introduction}
Experimental studies of charmonium states and their decay properties are important to test quantum Chromodynamics (QCD) models and QCD based calculations.
In the quark model, the $\chi_{cJ} (J = 0, 1, 2)$ mesons are identified as $^3P_J$ charmonium states.
Unlike the vector charmonium states $J/\psi$ and $\psi(3686)$, however, the $\chi_{cJ}$ mesons can not be directly produced in $e^+e^-$ collisions due to parity conservation, and our knowledge about their decays is relatively deficient. These $P$-wave charmonium mesons are produced abundantly via radiative $\psi(3686)$ decays, with branching fractions of about 9\%, thereby offering a good opportunity to study various $\chi_{cJ}$ decays.
Currently, theoretical studies indicate that the color octet mechanism (COM)~\cite{com} may substantially influence the decays of the $P$-wave charmonium states. However, some discrepancies between these theoretical calculations and experimental measurements have been reported in Refs.~\cite{ref::theroy1,ref::theroy2,ref::theroy3,ref::pdg2022}.
Therefore, intensive measurements of exclusive $\chi_{cJ}$ hadronic decays are highly desirable to understand the underlying $\chi_{cJ}$ decay dynamics.

\setlength{\parskip}{1ex}
In this paper we present the first observation and branching fraction measurements of $\chi_{cJ} \to 3(K^+K^-)$ by analyzing $(27.12\pm0.14)\times10^8$ $\psi(3686)$ events~\cite{ref::psip-num-inc} collected with the BESIII detector~\cite{ref::BesIII}.

\section{BESIII DETECTOR AND MONTE CARLO SIMULATION}
\label{sec:BES}
The BESIII detector~\cite{ref::BesIII} records symmetric $e^+e^-$ collisions provided by the BEPCII storage ring~\cite{ref::collider}
in the center-of-mass energy range from 2.0 to 4.95~GeV,
with a peak luminosity of $1 \times 10^{33}\;\text{cm}^{-2}\text{s}^{-1}$
achieved at $\sqrt{s} = 3.77\;\text{GeV}$.
The cylindrical core of the BESIII detector covers 93\% of the full solid angle and consists of a helium-based
 multilayer drift chamber~(MDC), a plastic scintillator time-of-flight
system~(TOF), and a CsI(Tl) electromagnetic calorimeter~(EMC),
which are all enclosed in a superconducting solenoidal magnet
providing a 1.0~T magnetic field.
The solenoid is supported by an
octagonal flux-return yoke with resistive plate counter muon
identification modules interleaved with steel.
The charged-particle momentum resolution at $1~{\rm GeV}/c$ is
$0.5\%$, and the
${\rm d}E/{\rm d}x$
resolution is $6\%$ for electrons
from Bhabha scattering. The EMC measures photon energies with a
resolution of $2.5\%$ ($5\%$) at $1$~GeV in the barrel (end cap)
region. The time resolution in the TOF barrel region is 68~ps, while
that in the end cap region was 110~ps.
The end-cap TOF system was upgraded in 2015 using multi-gap resistive plate chamber technology, providing a time resolution of 60~ps~\cite{Tof1,Tof2,Tof3}.

Simulated data samples produced with a {\sc
geant4}-based~\cite{Geant4} Monte Carlo (MC) package, which
includes the geometric description of the BESIII detector and the
detector response, are used to determine detection efficiencies
and to estimate backgrounds. The simulation models the beam
energy spread and initial state radiation (ISR) in the $e^+e^-$
annihilations with the generator {\sc
kkmc}~\cite{Jadach01}.
The inclusive MC sample includes the production of the
$\psi(3686)$ resonance, the ISR production of the $J/\psi$, and
the continuum processes incorporated in {\sc
kkmc}~\cite{Jadach01}.
All particle decays are modelled with {\sc
evtgen}~\cite{Lange01} using branching fractions
either taken from the
Particle Data Group (PDG)~\cite{ref::pdg2022}, when available,
or otherwise estimated with {\sc lundcharm}~\cite{Lundcharm00}.
Final state radiation~(FSR)
from charged final state particles is incorporated using the {\sc
photos} package~\cite{PHOTOS}.
An inclusive MC sample containing $2.7\times10^{9}$ generic $\psi(3686)$ decays is used to study background.
To account for the effect of intermediate
resonance structure on the efficiency, each of these
decays is modeled by the corresponding mixed signal
MC samples, in which the dominant decay modes containing resonances of $\phi$ are mixed with the phase-space (PHSP) signal MC samples. The mixing ratios are determined by examining the corresponding invariant mass as discussed in Section VI.

\section{EVENT SELECTION}
\label{sec:selection}

We reconstruct the events containing the charmonium
transitions $\psi(3686)\to\gamma\chic{J}$ followed by the hadronic
decays $\chic{J}\to 3(K^+K^-)$. The signal events are required to have at least six charged tracks and at least one photon candidate.

All charged tracks detected in the MDC are required to be within a polar angle ($\theta$) range of $|\rm{cos\theta}|<0.93$, where $\theta$ is defined with respect to the $z$-axis, which is the symmetry axis of the MDC. The distance of closest approach to the interaction point (IP)
must be less than 10\,cm
along the $z$-axis, $|V_{z}|$,
and less than 1\,cm
in the transverse plane, $|V_{xy}|$.
Particle identification~(PID) for charged tracks combines measurements of the energy deposited in the MDC~(d$E$/d$x$) and the flight time in the TOF to form likelihoods $\mathcal{L}(h)~(h=p,K,\pi)$ for each hadron $h$ hypothesis.
Tracks are identified as protons when the proton hypothesis has the greatest likelihood ($\mathcal{L}(p)>\mathcal{L}(K)$ and $\mathcal{L}(p)>\mathcal{L}(\pi)$), while charged kaons and pions are identified by comparing the likelihoods for the kaon and pion hypotheses, $\mathcal{L}(K)>\mathcal{L}(\pi)$ and $\mathcal{L}(\pi)>\mathcal{L}(K)$, respectively. Those with likelihood for kaon hypothesis greater than that for pion hypothesis are assigned to be kaon candidates.

Photon candidates are identified using showers in the EMC. The deposited energy of each shower must be more than 25~MeV in the barrel region ($|\cos \theta|< 0.80$) and more than 50~MeV in the end cap region ($0.86 <|\cos \theta|< 0.92$). To exclude showers that originate from charged tracks, the angle subtended by the EMC shower and the position of the closest charged track at the EMC must be greater than 10 degrees as measured from the IP. To suppress electronic noise and showers unrelated to the event, the difference between the EMC time and the event start time is required to be within [0, 700]\,ns.

\begin{figure}[htbp]
  \centering
  \includegraphics[width=\columnwidth]{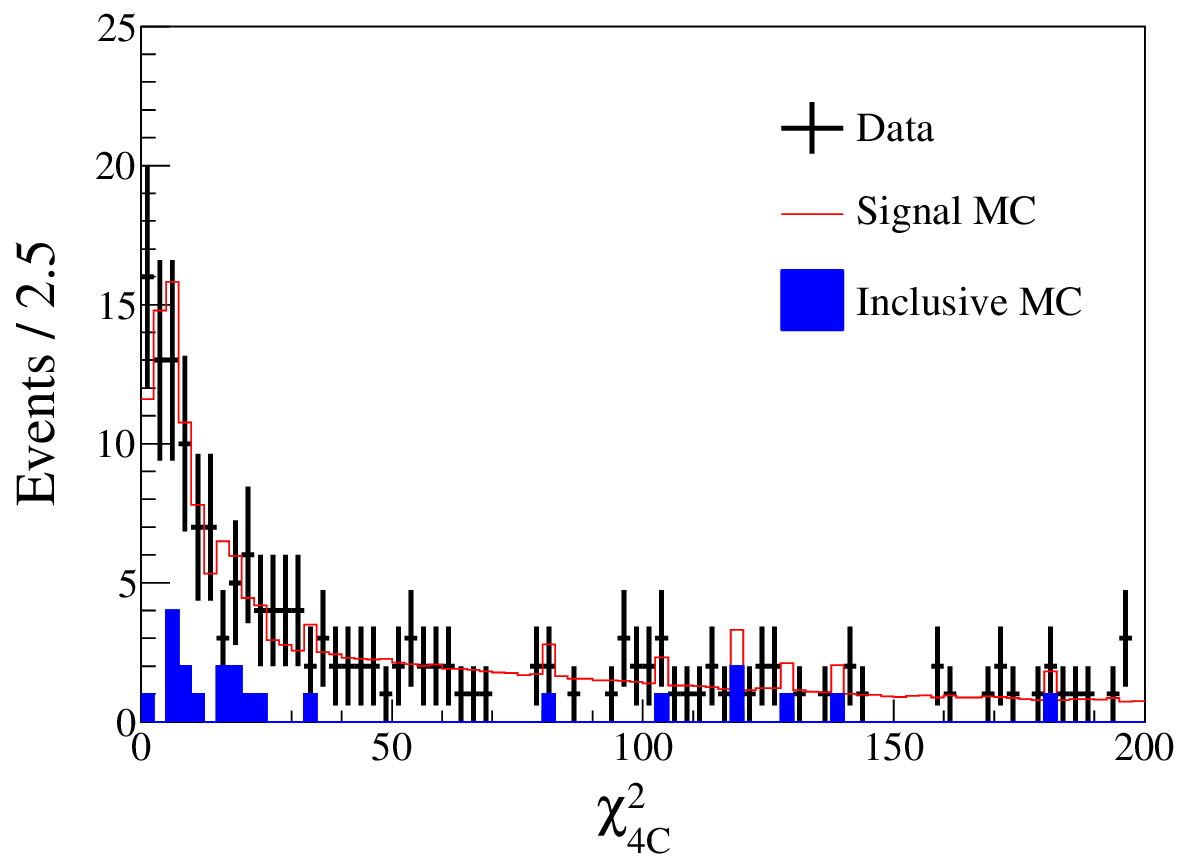}
  \caption{Distributions of $\chi^{2}_{\rm 4C}$ of the accepted candidate events. The dots with error bars are data, the red solid line is the signal MC sample that has been normalized to the data size, and the blue hatched histogram depicts the inclusive MC sample, which has also been normalized to the data size.}
 \label{tab:Kmfit}
\end{figure}

A four-momentum conservation constraint (4C) kinematic fit is applied to the events. In each event, if more than one combination survives, the one with the smallest $\chi_{\rm 4C}^{2}$ value of the 4C fit is retained.
Figure~\ref{tab:Kmfit} shows the $\chi^2_{\rm 4C}$ distributions of the accepted candidate events for data and MC samples.

The requirement on $\chi^2_{\rm 4C}$ is optimized with the Figure of Merit (FOM)
\begin{equation}
\mathrm{FOM} = \frac{\mathit{S}}{\sqrt{\mathit{S}+\mathit{B}}}.
\end{equation}
Here $S$ denotes the number of events from the signal MC sample, normalized according to the pre-measured branching fractions;
$B$ denotes the number of background events from the inclusive MC sample, normalized to the data size. After optimization, we choose $\chi^2_{\rm 4C}<50$ as the nominal requirement.

\section{BACKGROUND Analysis}
\label{sec:background}

The continuum data collected at $\sqrt s =$ 3.650 and 3.682 GeV, corresponding to an integrated luminosity of 800~pb$^{-1}$~\cite{lum}, are used to estimate the QED background. No event satisfies the same selection criteria applied to $\psi(3686)$ data.
Furthermore,
the inclusive MC sample is used to study all potential backgrounds from $\psi(3686)$ decays, and no event is observed in the $\chi_{cJ}$ signal regions.
Consequently, all peaking background components are treated as negligible in this analysis.

\section{Data analysis}

\begin{figure}[htbp]
  \centering
  \includegraphics[width=\columnwidth]{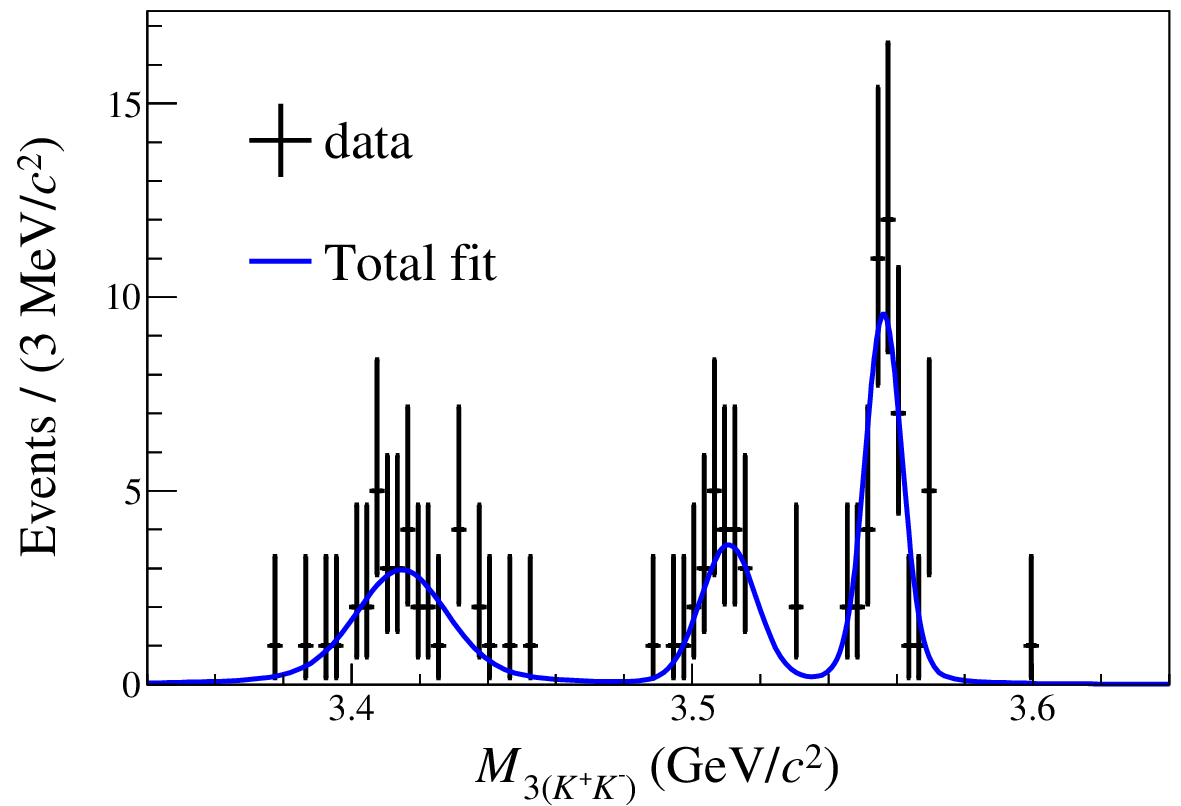}
  \caption{Fit to the $M_{3(K^+K^-)}$ distribution of the accepted candidate events. The points with error bars are data, the blue curve is the overall fit, and the red dashed line is the fitted background. }
  \label{fig:fit}
\end{figure}

 The distribution of the invariant mass of the $3(K^+K^-)$ combination, $M_{3(K^+K^-)}$, of the accepted candidate events is shown in Fig.~\ref{fig:fit}. Clear \chic{0}, \chic{1} and \chic{2} signals are observed.
The signal yields of $\chi_{cJ}\to 3(K^+K^-)$ are obtained from an unbinned maximum likelihood fit to this distribution.

In the fit, the signal shape of each $\chi_{cJ}$ is described by a Breit-Wigner functions convolved with a Gaussian. The widths and masses of Breit-Wigner functions are fixed to PDG averages~\cite{ref::pdg2022} for $\chi_{c0,1,2}$, respectively. The parameters of the Gaussian are floated.
 From this fit, the signal yields of $\chi_{c0}$, $\chi_{c1}$, and $\chi_{c2}$, $N_{\chi_{cJ}}^{\rm obs}$, are obtained to be $37.4\pm6.3$, $24.6\pm5.2$, and $46.3\pm7.0$, respectively. The statistical significances are estimated to be 9.5$\sigma$, $9.0\sigma$, and $13.7\sigma$ for \chic{0}, \chic{1}, and \chic{2} individually, which are determined by comparing the fit likelihood values separately with and without each $\chi_{cJ}$ signal component.

\section{Detection efficiency}

The efficiencies of detecting $\psi(3686)\to\gamma\chi_{cJ}$ with $\chi_{cJ}\to 3(K^+K^-)$
are determined with the mixed signal MC sample with fractions of the components of $\chi_{cJ}\to 2\phi (K^+K^-)$, $\chi_{cJ}\to\phi2(K^+K^-)$, and $\chi_{cJ}\to 3(K^+K^-)$ derived from a three-dimensional fit on the three $K^+K^-$ invariant mass spectra of the data events.
Table \ref{weight} shows the fractions of the sub-resonant decays.
The variations of these fractions are taken as systematic uncertainties.
The obtained detection efficiencies for
$\chi_{cJ}\to 3(K^+K^-)$ are $(13.3 \pm 0.1)\times 10^{-3}$, $(22.3 \pm 0.1)\times 10^{-3} $, and $(25.0 \pm 0.2)\times 10^{-3}$, respectively, including detector acceptance as well as reconstruction and selection efficiencies.

\begin{table}[htbp]
   \centering
    \caption{ The fractions of the sub-resonant decays for the mixed signal MC events.\label{tab:wei}}
\begin{tabular}{l|ccc}
 \hline
 \hline
&  $2\phi K^+ K^-$ & $\phi2(K^+ K^-)$ & $3(K^+K^-)$\\
\hline
$\chi_{c0} $ & $0.480^{+0.167}_{-0.151}$ & $0.038^{+0.306}_{-0.038}$ & $0.481^{+0.139}_{-0.158}$ \\
\hline
$\chi_{c1} $ &  $1.000^{+0.131}_{-0.005}$ & $0.000^{+0.125}_{-0.000}$ & $0.000^{+0.041}_{-0.000}$ \\
\hline
$\chi_{c2} $ &  $0.783^{+0.243}_{-0.180}$ & $0.217^{+0.179}_{-0.180}$ & $0.000^{+0.222}_{-0.000}$ \\
\hline
\hline
\end{tabular}
\label{weight}
\end{table}

\section{BRANCHING FRACTION}
\label{sec:mc}

For each decay
$\psi(3686)\to\gamma\chic{J}$, $\chic{J}\to 3(K^+K^-)$, about $10.8\times10^5$ signal MC events are generated
using a $1+\lambda\cos^2\theta$ distribution, where $\theta$ is the angle between the radiative photon and beam directions, and $\lambda=1,-1/3,1/13$ for $J=0,1,2$ in accordance with
the expectations for electric dipole transitions~\cite{ref::generate}.
Intrinsic width and mass values in PDG~\cite{ref::pdg2022} are used to simulate the \chic{J} states.

The product of branching fractions of $\psi(3686)\to \gamma\chi_{cJ}$ with $\chi_{cJ}\to 3(K^+K^-)$ is calculated as
\begin{equation}
\mathcal{B}_{\chi_{cJ}\to 3(K^+K^-)} \cdot  {\mathcal B}_{\psi(3686)\to \gamma\chi_{cJ}}=\frac{N^{\rm obs}_{\chi_{cJ}}}{N_{\psi(3686)}\cdot\epsilon},
\end{equation}
where $\epsilon$ is the detection efficiency and $N_{\psi(3686)}$ is the total number of $\psi(3686)$ events in data. Combining the branching fractions of $\psi(3686)\to\gamma\chi_{cJ}$ decays quoted from the PDG~\cite{ref::pdg2022}, the branching fractions of $\chi_{cJ} \to 3(K^+K^-)$ are determined. The obtained results are summarized in Table \ref{tab:Branching}.

\begin{table*}[htbp]\small
    \centering
    \caption{Signal yields in data, detection efficiencies, and branching fractions
    $\mathcal{B}({\psi(3686)\to\gamma\chi_{cJ}})\cdot \mathcal{B}({\chi_{cJ}\to 3(K^+K^-)})$, and $\mathcal{B}({\chi_{cJ}\to 3(K^+K^-)})$. Except for $\mathcal B_{\psi(3686)\to\gamma\chi_{cJ}}$, the uncertainties are statistical only.}

     \begin{tabular}{l  r@{ $\pm$ }c r@{ $\pm$ }c r@{ $\pm$ }c}
   \hline
   \hline
&\multicolumn{2}{c}{$\chi_{c0}$}  &\multicolumn{2}{c}{$\chi_{c1}$} & \multicolumn{2}{c}{$\chi_{c2}$} \\
\hline
  $N_{\rm obs}$ & 37.7&6.2  & 24.9&5.1  & 46.4&7.0                 \\
  $\epsilon~(\times 10^{-3})$ & 13.3&0.1          & 22.3&0.1          & 25.0&0.2          \\
  $\mathcal{B}_{\psi(3686)\to \gamma\chi_{cJ}}\cdot \mathcal{B}_{\chi_{cJ}\to 3(K^+K^-)}~(\times 10^{-7})$ &10.5&1.8& 4.1&0.9 & 6.8&1.1 \\

 $\mathcal{B}_{\chi_{cJ}\to 3(K^+K^-)}~(\times 10^{-6})$   &10.7&1.8& 4.2&0.9& 7.2&1.1\\
   \hline
   \hline

\end{tabular}
\label{tab:Branching}

\end{table*}

\section{SYSTEMATIC UNCERTAINTY}
\label{sec:systematics}

The systematic uncertainties in the branching fraction measurements originate from several sources, as summarized in Table~\ref{tab:systematics}. They are estimated and discussed below.
\begin{table}[htbp]
  \caption{Relative systematic uncertainties in the branching fraction measurements~(\%). The last item is the systematic uncertainty of the introduced reference.}
  \begin{tabular} {lrrr}
  \hline\hline
  Source  & $\chi_{c0}$& $\chi_{c1}$& $\chi_{c2}$\\
        \hline
  $N_{\psi(3686)}$         & 0.5 & 0.5 & 0.5\\
  $K^\pm$ tracking         & 6.0 & 6.0 & 6.0\\
  $K^\pm$ PID              & 6.0 & 6.0 & 6.0\\
  $\gamma$ selection       & 1.0 & 1.0 & 1.0\\
  Fractions of different sub-processes	&    3.3     &      0.8     &    2.4 \\
 $M_{3(K^+K^-)}$ fit       & 3.3 & 7.0 & 5.2\\
 4C kinematic fit          & 3.0 & 3.0 &3.0 \\
  MC statistics            & 1.6 & 1.2 & 1.1\\
  Sum                      & 10.3&11.5&10.8\\
  \hline
 $\mathcal{B}(\psi(3686) \to \gamma\chi_{cJ})$ & 2.0& 2.4& 2.0\\
  Total                   & 10.5 & 11.7 & 11.0\\
   \hline
   \hline
  \end{tabular}
  \label{tab:systematics}
\end{table}

The total number of $\psi(3686)$ events in data has been measured to be $N_{\psi(3686)}=(27.12\pm0.14)\times10^8$ with the inclusive
hadronic data sample, as described in Ref.~\cite{ref::psip-num-inc}. The uncertainty of $N_{\rm \psi(3686)}$ is 0.5\%.

The systematic uncertainty of the $K^\pm$ tracking or PID efficiencies is  assigned as 1.0\% per $K^\pm$\cite{ref::tracking}, which is estimated with the control samples of $J/\psi\to K^{*}\bar{K}$.

The systematic uncertainty in the photon detection is assumed to be 1.0\% per photon with the control sample $J/\psi\to\pi^+\pi^-\pi^0$~\cite{ref::gamma-recon}.

To estimate the systematic uncertainties of the MC model for the $\chi_{cJ}\to 3(K^+K^-)$ decays, we compare our nominal efficiencies with those determined from the signal MC events after varying $\pm 1$ standard deviation of the relative fractions of the sub-resonant decays, including $\chi_{cJ}\to 2\phi K^+K^-$, $\chi_{cJ}\to \phi 2(K^+K^-)$,
and $\chi_{cJ}\to 3(K^+K^-)$. The relative changes of efficiencies, which are 3.3\%, 0.8\%, and 2.4\% for \chic{0}, \chic{1}, and \chic{2} decays respectively, are assigned as the corresponding systematic uncertainties.

The systematic uncertainty of the fit to the $M_{3(K^+K^-)}$ spectrum includes three parts:
\begin{itemize}
  \item The first is the background shape estimated by allowing a slope in the background. The changes of the fitted signal yields, 1.4$\%$ for $\chi_{c0}$, 6.5$\%$ for $\chi_{c1}$, 4.4$\%$ for $\chi_{c2}$, are taken as the corresponding systematic uncertainties.

  \item 	
  The second is from the signal shape, which is estimated by varying the width of the $\chi_{cJ}$ state by $\pm 1$ standard deviation. The change of the fitted signal yield of each decay is negligible.
  \item
  The third is due to the fit range estimated with alternative ranges of $[3.225, 3.635]$, $[3.225, 3.615]$,
$[3.215, 3.625]$, $[3.235, 3.625]$, $[3.225,3.625]$~GeV/$c^2$. The maximum changes of the fitted signal yields, 3.0$\%$ for $\chi_{c0}$, 2.8$\%$ for $\chi_{c1}$, and 2.8$\%$ for $\chi_{c2}$ are taken as the corresponding systematic uncertainties.
  \end{itemize}

The systematic uncertainty resulting from the $M_{3(K^+K^-)}$ fit is determined be 3.3$\%$ for $\chi_{c0}$, 7.0$\%$ for $\chi_{c1}$, and 5.2$\%$ for $\chi_{c2}$, when combining these three uncertainties in quadrature.

The systematic uncertainty of the 4C kinematic fit comes from the inconsistency between the data and MC simulation of the track-helix parameters. We make helix parameter corrections to take the difference between the efficiencies with and without the corrections as the systematic uncertainty. The systematic uncertainties of the 4C kinematic fits are obtained to be 3\% for all decays $\chi_{cJ} \to (K^+K^-)$ ($J=0,1,2$).

The systematic uncertainties due to the statistics of the MC samples
are 1.6\%, 1.2\%, and 1.1\% for $\chi_{c0}$, $\chi_{c1}$, and $\chi_{c2}$ decays, respectively.

The systematic uncertainties from the branching fractions of $\psi(3686)\to\gamma\chic{J}$ decays quoted from the PDG~\cite{ref::pdg2022} are 2.0\%, 2.4\%, and 2.0\% for $\chi_{c0}$, $\chi_{c1}$, and $\chi_{c2}$ decays, respectively.

We assume that all systematic uncertainties are independent and
combine them in quadrature to obtain the total systematic uncertainty for each decay.

\section{Summary}

By analyzing $(27.12\pm0.14)\times10^8$ $\psi(3686)$ events with the BESIII detector,
the product branching fractions of $\psi(3686)\to\gamma\chi_{cJ}$, $\chi_{cJ}\to 3(K^+K^-)$ are determined to be $\mathcal{B}_{\psi(3686)\to\gamma\chi_{c0}}\cdot\mathcal{B}_{\chic{0}\to 3(K^+K^-)}=$$(10.5\pm1.8)$$\times10^{-5}$, $\mathcal{B}_{\psi(3686)\to\gamma\chi_{c1}}\cdot\mathcal{B}_{\chic{1}\to 3(K^+K^-)}=$$(4.1\pm0.9)$$\times10^{-5}$, and $\mathcal{B}_{\psi(3686)\to\gamma\chi_{c2}}\cdot\mathcal{B}_{\chic{2}\to 3(K^+K^-)}=$$(6.8\pm1.1)$$\times10^{-5}$, where the uncertainties are statistical. The decays of $\chi_{cJ} \to 3(K^+K^-)$ are observed for the first time with statistical significances of 8.2$\sigma$, 8.1$\sigma$, and 12.4$\sigma$, respectively. We measure the branching fractions of
$\chic{J}\to 3(K^+K^-)$ to be $\mathcal{B}_{\chic{0}\to 3(K^+K^-)}=$$(10.7\pm1.8\pm1.1)$$\times10^{-6}$,
$\mathcal{B}_{\chic{1}\to 3(K^+K^-)}$=$(4.2\pm0.9\pm0.5)$$\times10^{-6}$, $\mathcal{B}_{\chic{2}\to 3(K^+K^-)}$=$(7.2\pm1.1\pm0.8)$$\times10^{-6}$, where the first uncertainties are statistical and the second systematic.
These results offer additional data for understanding of the decay mechanisms of $\chi_{cJ}$ states.

\section{ACKNOWLEDGMENTS}

The BESIII Collaboration thanks the staff of BEPCII and the IHEP computing center for their strong support. This work is supported in part by National Key R\&D Program of China under Contracts Nos. 2020YFA0406300, 2020YFA0406400; National Natural Science Foundation of China (NSFC) under Contracts Nos. 11635010, 11735014, 11835012, 11935015, 11935016, 11935018, 11961141012, 12025502, 12035009, 12035013, 12061131003, 12192260, 12192261, 12192262, 12192263, 12192264, 12192265, 12221005, 12225509, 12235017; the Chinese Academy of Sciences (CAS) Large-Scale Scientific Facility Program; the CAS Center for Excellence in Particle Physics (CCEPP); Joint Large-Scale Scientific Facility Funds of the NSFC and CAS under Contract No. U1832207; CAS Key Research Program of Frontier Sciences under Contracts Nos. QYZDJ-SSW-SLH003, QYZDJ-SSW-SLH040; 100 Talents Program of CAS; The Institute of Nuclear and Particle Physics (INPAC) and Shanghai Key Laboratory for Particle Physics and Cosmology; European Union's Horizon 2020 research and innovation programme under Marie Sklodowska-Curie grant agreement under Contract No. 894790; German Research Foundation DFG under Contracts Nos. 455635585, Collaborative Research Center CRC 1044, FOR5327, GRK 2149; Istituto Nazionale di Fisica Nucleare, Italy; Ministry of Development of Turkey under Contract No. DPT2006K-120470; National Research Foundation of Korea under Contract No. NRF-2022R1A2C1092335; National Science and Technology fund of Mongolia; National Science Research and Innovation Fund (NSRF) via the Program Management Unit for Human Resources \& Institutional Development, Research and Innovation of Thailand under Contract No. B16F640076; Polish National Science Centre under Contract No. 2019/35/O/ST2/02907; The Swedish Research Council; U. S. Department of Energy under Contract No. DE-FG02-05ER41374.

\end{document}